\documentclass[a4paper,11pt]{article}
\pdfoutput=1 \usepackage{jheppub_2} 
\usepackage{booktabs}
\usepackage[T1]{fontenc} 
\usepackage{graphicx, multirow,soul,url,amsmath,amsfonts,amssymb,mathrsfs,amsfonts}
\usepackage[all]{hypcap}
\usepackage{url}
\usepackage{bbm}
\usepackage{changepage}
\usepackage{cancel}
\usepackage{sidecap}
\usepackage{graphicx}
\usepackage{slashed}
\usepackage[dvipsnames]{xcolor}
\usepackage{multicol, blindtext}
\usepackage[section]{placeins}
\usepackage[version=3]{mhchem}
\usepackage{caption}
\expandafter\def\csname ver@subfig.sty\endcsname{}
\usepackage{svg}
\usepackage{lipsum}
\usepackage{hyperref}
\usepackage{float}
\usepackage{mattens}
\usepackage{mathtools}
\usepackage{footnote}
\usepackage{amsmath,amssymb,amsthm,amsxtra,overpic,bbm,bm,epsfig}
\usepackage[splitrule]{footmisc}
\usepackage{bbold}
\usepackage{amsmath}
\usepackage{amsbsy}
\usepackage{lipsum}
\usepackage[export]{adjustbox}
\usepackage{units}

\newcommand{\equaref}[1]{Eq.~(\ref{#1})}

\newcommand{\figref}[1]{Fig.~\ref{#1}}

\newcommand{\secref}[1]{Section~\ref{#1}}
\newcommand{\secrefs}[2]{Sections~\ref{#1}~and~\ref{#2}}
\newcommand{\appref}[1]{Appendix~\ref{#1}}
\newcommand{\tabref}[1]{Table~\ref{#1}}

\newcommand{\bq}{\begin{eqnarray}}
\newcommand{\nq}{\end{eqnarray}}

\usepackage{subfig}

\title{\bf Light neutrino masses from gravitational condensation: the Schwinger-Dyson approach}
\author[a]{Gabriela Barenboim,}
\affiliation[a]{Departament de Fisica Teorica and IFIC, Universitat de Valencia, 46100 Burjassot, Spain.}
\author[b]{Jessica Turner}
\affiliation[b]{Institute for Particle Physics Phenomenology, Durham University, South Road, Durham, U.K.}
\author[c]{and Ye-Ling Zhou}
\affiliation[c]{School of Physics and Astronomy, University of Southampton, Southampton, SO17 1BJ, U.K.}

\emailAdd{gabriela.barenboim@uv.es}
\emailAdd{jessica.turner@durham.ac.uk}
\emailAdd{ye-ling.zhou@soton.ac.uk}

\abstract{
In this work we demonstrate that non-zero neutrino masses can be generated from gravitational interactions. 
We solve the Schwinger-Dyson equations to find a non-trivial vacuum thereby determining the neutrino condensate scale and the number of new particle degrees of freedom required
for gravitationally induced dynamical chiral symmetry breaking. We show for minimal beyond the Standard Model particle content, the scale of the condensation occurs
close to the Planck scale.
}

\preprint{
\begin{flushleft}{
IFIC/19-41\\
FERMILAB-PUB-19-461-T}
\end{flushleft}
}


\begin{document}

\thispagestyle{empty}
\def\thefootnote{\fnsymbol{footnote}}
\setcounter{footnote}{1}

\setcounter{page}{0}
\maketitle
\vspace{-1cm}
\flushbottom

\def\thefootnote{\arabic{footnote}}
\setcounter{footnote}{0}

\section{Introduction}\label{sec:intro}
Neutrinos are unique amongst the Standard Model (SM) fermions in their mass's tininess, the weakness of their interactions and their capacity to be their own anti-particles. Such features suggest neutrinos acquire their mass differently from the quarks and charged leptons.  Many such  mass models assume neutrinos are Majorana particles and the most prolifically studied  are the seesaw mechanisms \cite{Mohapatra:1979ia,GellMann:1980vs,Yanagida:1979as,Minkowski:1977sc,Magg:1980ut,Lazarides:1980nt,Wetterich:1981bx,Mohapatra:1980yp,Schechter:1980gr,Schechter:1981cv}. 
Typically the masses of the new particles required to complete the 
lepton-number violating Weinberg operator are larger than the 
 electroweak scale. In addition to tree-level completions of the  Weinberg operator,  radiative mass models can explain small neutrino masses with TeV-scale new physics \cite{Zee:1980ai,Cheng:1980qt,Petcov:1982en,Babu:1988ki}. 
Moreover,  explanations of light neutrino masses from extra-dimensions \cite{ArkaniHamed:1998vp,Dienes:1998sb} and  string theory \cite{Mohapatra:1986bd} provide alternative possibilities (see Ref. \cite{King:2013eh} for an extensive overview of models of neutrino masses and mixing).

Neutrino masses emerging from gravitational effects were first discussed in \cite{Akhmedov:1992hh} where Planck suppressed higher-dimensional operators
induced neutrino masses. Furthermore,  the possibility of neutrino masses emerging from a  gravitationally triggered condensate has been studied in various contexts:
in \cite{Barenboim:2010db} it was shown that enhanced gravitational interactions could trigger the formation of a right handed neutrino condensate which induces dynamical symmetry breaking and thereby generates a Majorana mass for the right handed neutrino. From this, the light neutrino masses are generated via the type-I seesaw mechanism.
An advantage of such an approach is that the strongly coupled right handed neutrino condensate can drive inflation \cite{Barenboim:2008ds}.
A more direct explanation for light neutrino masses was proposed in \cite{Dvali:2016uhn}. In that work it was postulated that gravitational instantons could induce a low-scale ($\sim 100$ meV) neutrino condensation which can give rise to light neutrino masses \cite{Dvali:2016uhn};
the phenomenology of which has been studied in depth \cite{Funcke:2019grs}.

In this work, we use the Schwinger-Dyson equations to demonstrate that an enhanced gravitational attraction can trigger the formation of an active neutrino condensate which induces dynamical symmetry breaking. 
We treat gravity as an effective quantum field theory in the spirit of \cite{Donoghue:1994dn}. 
With minimal assumptions, we show that a non-trivial vacuum can be achieved and find that the phase transition scale is  close to the Planck scale. However,  new particle degrees of freedom are required to provide finite support to the condensate.
Neutrinos remain free particles below the scale of condensation (similarly to \cite{Dvali:2016uhn}) in analogy to the
Nambu-Jona-Lasinio (NJL) model \cite{Nambu:1961tp} where  the constituent
fermions are free (unconfined). Furthermore, due to gravity's democratic nature, it will provide a small mass for all fermions. 

The work presented in this paper is structured as follows: in \secref{sec:SDEs} we review the Schwinger-Dyson equations (see Ref. \cite{Roberts:1994dr} for an in-depth discussion of Schwinger-Dyson methods in QCD and QED) and discuss the leading order diagram which contributes to gravitationally induced neutrino chiral symmetry breaking. We find that chiral symmetry is preserved if the bare graviton propagator is used. Consequently, we apply the dressed graviton propagator, which is discussed in detail in \secref{sec:gravprop}. Further, in this section, we introduce the pertinent parameters upon which the neutrino masses depend: the condensate scale $\Lambda$ and two quantities which parametrise the particle content, $A$ and $B$. We elucidate the challenges of finding the chiral breaking vacuum and present two solutions to the Schwinger-Dyson equations in \secref{sec:extra} and \secref{sec:const} respectively. Finally, we summarise and make concluding remarks in \secref{sec:discussion}.

\section{The Schwinger-Dyson equation}\label{sec:SDEs}
The Schwinger-Dyson equations (SDE) are an infinite tower of integral coupled equations which relate the Green functions of a theory
to each other. From this set of coupled equations, all observables of the theory can be calculated. We use the SDE
as a tool to demonstrate that active neutrinos, which we assume have zero bare mass, can condense via their gravitational
interactions and thereby undergo dynamical chiral symmetry breaking. This phenomenon is ultimately non-perturbative, and the SDE provides a method to derive the neutrino gap equation. 

The leading order propagator for a massless fermion  is simply $S_{F}=i/\slashed{p}$. The full propagator will receive self-energy corrections 
which modify its form in the following way
\begin{equation}\label{eq:SDE1}
S'_F(p) = \frac{i}{p\!\!\!\slash - \Sigma(p)} = \frac{i}{\alpha(p^2)p\!\!\!\slash - \beta(p^2)} \,,
\end{equation}
where $\alpha(p^2)$ and $\beta(p^2)$ are determined by the relevant self-energy correction, $\Sigma(p)$, and the dynamically
induced mass of the fermion is $m_F=\beta(p^2)/\alpha(p^2)$. From \equaref{eq:SDE1}, we find the propagator consists of two parts: the Dirac odd component, which is the scalar function $\alpha(p^2)$, and the
Dirac even  part which is parametrised by  $\beta(p^2)$. Using the appropriate Dirac trace we find the correlation of these functions with the self-energy correction to be
\begin{equation} 
\label{eq:alpha_beta}
\alpha(p^2) = 1 - \frac{1}{4p^2}{\rm tr}(p\!\!\!\slash \Sigma(p)) \,,\quad
\beta(p^2) = \frac{1}{4}{\rm tr}(\Sigma(p)) \,. 
\end{equation}

\begin{figure}[t!]
\centering
\includegraphics[width=0.45\textwidth]{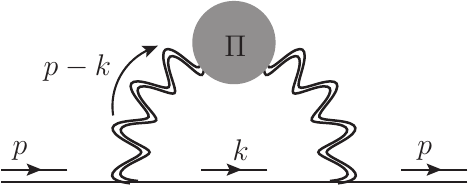}
\caption{The self-energy correction to the neutrino propagator. The graviton (represented by the double wavy line) is dressed with the vacuum polarisation (indicated by $\Pi$) and  external arrows show the momentum flow.} \label{fig:neutprop}
\end{figure}

The leading gravitational self-energy correction, as shown in \figref{fig:neutprop}, to the fermion propagator is given by
\begin{equation}\label{eq:SDE2}
-i \Sigma(p) = \int \frac{d^4 k}{(2\pi)^4} \tau_1^{\mu\nu}(p,-k) S^\prime_F(k) G^\prime_{\mu\nu\rho\sigma}(p-k) \tau_1^{\rho\sigma}(k,-p) \,,
\end{equation}
where $\tau_1$ is the fermion-fermion-graviton vertex, $S^\prime_F$ is the modified fermion propagator and $G^\prime$ is the dressed
graviton propagator. 
Using the graviton Feynman rules and substituting them into 
 \equaref{eq:SDE2},  $\alpha(p^2)$ and $\beta(p^2)$ may be written as
\begin{equation} \label{eq:alpha_beta}
\begin{aligned}
&\alpha(p^2) = 1 +\frac{1}{4p^2} \int [dk]{\rm tr}\left[p\!\!\!\slash \tau_1^{\mu\nu}(p,-k)f(k) G'_{\mu\nu\rho\sigma}(p-k) \tau_1^{\rho\sigma}(k,-p) \right] \,, \\
&\beta(p^2) = - \frac{1}{4}  \int [dk]{\rm tr}\left[ \tau_1^{\mu\nu}(p,-k)f(k) G'_{\mu\nu\rho\sigma}(p-k) \tau_1^{\rho\sigma}(k,-p) \right] \,.
\end{aligned}
\end{equation}
where 
\begin{equation}
f(k)= \frac{\alpha(k^2)k\!\!\!\slash + \beta(k^2)}{\alpha^2(k^2)k^2 - \beta^2(k^2)}\quad \text{and} \quad [dk] =\frac{d^4 k}{(2\pi)^4} 
\end{equation}
and we replace the dressed graviton propagator, $G'_{\mu\nu\rho\sigma}(p-k)$,  by its tree-level counterpart, $G_{\mu\nu\rho\sigma}(p-k)$,  we obtain the leading order contribution of $\alpha(p^2)$

\begin{equation} \label{eq:alpha_1}
\begin{aligned}
\alpha(p^2) &= 1 -\\
& i 2\pi G \int [dk]f(k)
\frac{ \left[ 2(k\cdot p)^2+4k^2 p^2+3k\cdot p(k^2+p^2) \right]  }{p^2(p-k)^2} \,,\\
\beta(p^2) & = 0\,.
\end{aligned}
\end{equation}
Remarkably, the leading gravitationally induced correction preserves chiral symmetry as  $\beta(p^2)$ is exactly zero for all momentum values. The dynamical breaking of  chiral symmetry manifests by dressing 
the  graviton propagator with matter fields and the graviton itself
at the one-loop order. 
More specifically,  we perform the following replacement

\begin{equation}
\begin{aligned}
G'_{\mu\nu\rho\sigma}(p-k) &\to \, G_{\mu\nu\rho\sigma}(p-k) \\
&+ G_{\mu\nu\alpha\beta}(p-k) \Pi^{\alpha\beta,\gamma\delta}(p-k) G_{\rho\sigma\gamma\delta}(p-k) \,,
\end{aligned}
\end{equation}
where $\Pi^{\alpha\beta,\gamma\delta}(p-k)$ are the vacuum polarisation diagrams as shown in \figref{fig:vacpol}.


\section{Graviton self-energy form factor}\label{sec:gravprop}
The vacuum polarisation tensor, $\Pi^{\alpha\beta,\gamma\delta}(q)$, which corrects the tree-level graviton propagator is a fourth rank tensor in $SO(1,3)$. This tensor can be constructed from   $\eta^{\alpha\beta}$ and $p^\alpha$ and may be written as a linear combination of five independent  tensors of rank four:
\begin{equation}
\begin{aligned}
&\Pi^{\alpha\beta,\gamma\delta}(q) 
= F_1(q^2) q^\alpha q^\beta q^\gamma q^\delta
+ F_2(q^2) \eta^{\alpha\beta} \eta^{\gamma\delta} \,\\
&+ F_3(q^2) (\eta^{\alpha\gamma} \eta^{\beta\delta} + \eta^{\alpha\delta} \eta^{\beta\gamma})
+ F_4(q^2) (q^\alpha q^\beta \eta^{\gamma\delta} + q^\gamma q^\delta \eta^{\alpha\beta}) \,\\
&+ F_5(q^2) (q^\alpha q^\gamma \eta^{\beta\delta} + q^\alpha q^\delta \eta^{\beta\gamma} + q^\beta q^\gamma \eta^{\alpha\delta} + q^\beta q^\delta \eta^{\alpha\gamma}) \,.
\end{aligned}
\end{equation}
The above vacuum polarisation expression is invariant under permutations $\alpha \leftrightarrow \beta$, $\gamma \leftrightarrow \delta$, as well as $\alpha\beta \leftrightarrow \gamma\delta$ and $F_{i}$ (where $i \in 1, 2 ..,5 $) are a set of form factors. 
The form factors are not independent of each other as the 
vacuum polarisation of the graviton must satisfy the Ward identity, $p_\alpha\Pi^{\alpha\beta,\gamma\delta}(q) = 0$, which leads to three constraints on the five tensors
\begin{equation}
\begin{aligned}
q^2 F_1 + F_4 +F_5 &=0\,,\\ 
F_2+q^2 F_4&=0\,,\\ 
F_3 + q^2 F_5&=0 \,.
\end{aligned}
\end{equation}

As we include up to the  one-loop correction to the graviton propagator we parametrise two of these
form factors as
\begin{equation}\label{eq:con3}
\begin{aligned}
F_4(q^2) &= a_1\, q^2\log \left[ \frac{\mu^2}{-q^2} \right] \,, \\
F_5(q^2) &= a_2\, q^2\log \left[ \frac{\mu^2}{-q^2} \right] \,, 
\end{aligned}
\end{equation}
where $\log[\mu^2/(-q^2)]$ comes from  the one-loop integration.
Using the three constraints of \equaref{eq:con3} we derive the following relations
\begin{equation}
\begin{aligned}
F_1(q^2) &= ( a_1 + 2 a_2 )\, \log \left[ \frac{\mu^2}{-q^2} \right] \,, \\
F_2(q^2) &= -a_1\, (q^2)^2 \log \left[ \frac{\mu^2}{-q^2} \right] \,, \\
F_3(q^2) &= -a_2\, (q^2)^2 \log \left[ \frac{\mu^2}{-q^2} \right] \,.
\end{aligned}
\end{equation}
Using these constraints, the vacuum polarisations may  be parametrised as
\vspace{0.5cm}

\begin{equation} \label{eq:form_factor}
\begin{aligned}
& \Pi^{\alpha\beta,\gamma\delta}(q) = a_1 ( q^\alpha q^\beta - \eta^{\alpha\beta} q^2 ) ( q^\gamma q^\delta - \eta^{\gamma\delta} q^2 )\log\left[\frac{\mu^2}{-q^2}\right] \\
&+ a_2 \left[ ( q^\alpha q^\gamma - \eta^{\alpha\gamma} q^2 ) ( q^\beta q^\delta - \eta^{\beta\delta} q^2 ) \right] \log\left[\frac{\mu^2}{-q^2}\right]\\
& + a_2 \left[ ( q^\alpha q^\delta - \eta^{\alpha\delta} q^2 ) ( q^\beta q^\gamma - \eta^{\beta\gamma} q^2 ) \right] \log\left[\frac{\mu^2}{-q^2}\right]\,.
\end{aligned}
\end{equation}

Contracting $\Pi^{\alpha\beta,\gamma\delta}(q^2)$ with $\eta_{\alpha\beta}\eta_{\gamma\delta}$ and $\eta_{\alpha\gamma}\eta_{\beta\delta}$ respectively, we obtain
\begin{equation}
\begin{aligned}
\Pi_1 &\equiv \eta_{\alpha\beta} \eta_{\gamma\delta}\Pi^{\alpha\beta,\gamma\delta} = (9 a_1 + 6 a_2) (q^2)^2 \log\left[\frac{\mu^2}{-q^2}\right] \,, \\
\Pi_2 &\equiv \eta_{\alpha\gamma}\eta_{\beta\delta}\Pi^{\alpha\beta,\gamma\delta} = (3 a_1 + 12 a_2) (q^2)^2 \log\left[\frac{\mu^2}{-q^2}\right] \,.
\end{aligned}
\end{equation}

\begin{figure*}[t!]
\centering
\includegraphics[width=0.85\textwidth]{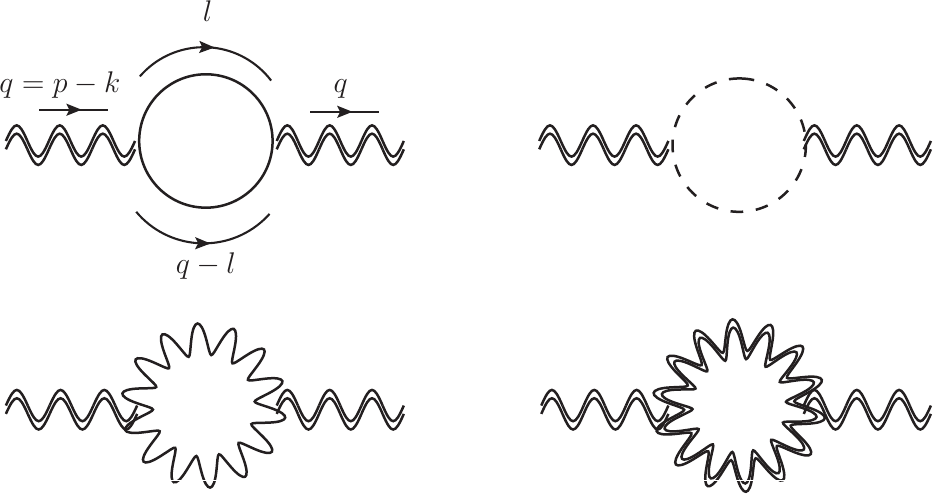}
\caption{The set of vacuum polarisations, $\Pi$, which modify the bare graviton propagator is represented by the double wavy lines. The solid line (dotted) indicates a  Dirac fermion (minimal scalar) field and the wavy (double wavy)
indicates a gauge boson (graviton). The external arrows indicate the momentum flow.}\label{fig:vacpol}
\end{figure*}
For any loops contributing to the vacuum polarisation once we calculate the Lorentz-invariants quantities $\Pi_1$ and $\Pi_2$ from the loop integration, we obtain $a_1$ and $a_2$, from which we derive the self-energy $\Pi^{\alpha\beta,\gamma\delta}$.\footnote{The above method exploits the transversality of the graviton self-energy in order to calculate the vacuum polarisations. 
The ``brute force''  method can be found in  \appref{sec:vacpolapp}.}

The vacuum polarisations used in this work match those calculated in \cite{Barenboim:2010db} where the vacuum polarisations to the graviton
from the minimal scalar, fermion and gauge bosons were calculated. The vacuum polarisations from the graviton contribution, along with the ghost contribution, were initially calculated in
\cite{tHooft:1974toh} and discussed in \cite{Donoghue:1994dn,BjerrumBohr:2004mz}. 

Given the  graviton Feynman rules for interactions with  a minimal scalar (ms), Dirac fermion (df), conformal scalar (cs), gauge boson (gb) and graviton (gr) as provided in \appref{sec:FR}, we obtain the values of $a_1$ and $a_2$  as shown in \tabref{tab:self-energy}. By contracting $\Pi^{\alpha\beta,\gamma\delta}$  with the tree-level propagators $G_{\mu\nu\alpha\beta}(p-k)$ and $G_{\rho\sigma,\gamma\delta}(p-k)$, we determine the dressed graviton propagator $G^\prime_{\mu\nu\alpha\beta}(p-k)$ which we substitute
 into \equaref{eq:alpha_beta} to find 
 
\begin{equation} \label{eq:beta_1}
\begin{aligned}
\beta(p^2) = i 8 G^2 \int[dk] f(k)
&\left[ A (k+p)^2 - B \frac{\left(p^2-k^2\right)^2}{8 (p-k)^2} \right]\\
& \log \left[\frac{\mu^2}{-(p-k)^2} \right] \,,
\end{aligned}
\end{equation}

where $\mu$ is the renormalisation mass which in principle is arbitrary. The degrees of freedom running in the loop diagrams of \figref{fig:vacpol} are constants given by
\begin{equation}
\begin{aligned}
A &= \frac{5}{16} \sum_p \left(5 a_1^{p}+6 a_2^{p}\right) N_{p} \,,\\
B &= \frac{1}{2} \sum_p \left(2 a_1^{p}+3 a_2^{p}\right) N_{p} \,,
\end{aligned}
\end{equation}
where $p$ is the index for the particle type (ms, df, gb, cs, gr) and $N^p$ is the number of each type of particle. By taking the values of $a_1^{p} $ and $a_2^{p} $ in \tabref{tab:self-energy} and fixing the degree of freedom for graviton to be one, we recover the result in Ref.~\cite{Barenboim:2010db}:
\footnote{ The overall structure of A and B is the same in this work and that of Ref.~\cite{Barenboim:2010db} up to a global factor of eight. This discrepancy stems from an error in Ref.~\cite{Barenboim:2010db}.
We note that the SM does not contain any conformal scalars and throughout this work we set Ncs = 0. For completeness we provide the expressions for A and B and associated coefficients in Table 1.}
\begin{equation}
\begin{aligned}
A &= \frac{27/2 N_{\rm ms} + 6 N_{\rm df} + 12 N_{\rm gb} + N_{\rm cs} + 267 N_{\rm gr}}{288} \,,\\
B &= \frac{9 N_{\rm ms} + 6 N_{\rm df} + 12 N_{\rm gb} + N_{\rm cs} + 186 N_{\rm gr}}{288} \,.
\end{aligned}
\end{equation}
The Standard Model  has a large number of degrees of freedom: 12 gauge bosons; 48 chiral fermions and four Higgs scalars.  As such the SM values of these parameters are  $A=2.61$ 
and $B=2.27$. However, it is possible there are many more new degrees of freedom at higher energy scales and there are a plethora of theories
which consider non-minimal particle content. 
For example, in the Minimal Supersymmetric Standard Model, these parameters are enlarged such that $A=5.19$ and $B=4.10$.  There are other theories with an even richer particle spectrum,
for instance the Scalar Democracy as outlined in  \cite{Hill:2019ldq}  predicts the existence of 1176 Higgs doublets as a dynamical explanation for the observed fermion mass hierarchy and mixing.  In such a theory, $A=223.15$ and $B=149.31$. Moreover, theories which ensure the asymptotic safety of the Standard Model \cite{Abel:2018fls} predict similar values of $A$ and $B$  to the work mentioned above. 
 Likewise, a large number of copies ($\sim 10^{32}$) of the SM was used to explain the origin and nature of dark matter \cite{Dvali:2009fw} and correspond to very large $A\sim B \approx 2\times 10^{32}$. 
\begin{table}[t!]
\centering 
\begin{tabular}{lcc}\hline\hline
particle in the loop & ~~$a_1/\frac{G}{\pi}$~~ & ~~$a_2/\frac{G}{\pi}$~~ \\\hline
minimal scalar & $\frac{1}{40}$ & $\frac{1}{240}$ \\
Dirac fermion & $-\frac{1}{60}$ & $\frac{1}{40}$ \\
gauge boson & $-\frac{1}{30}$ & $\frac{1}{20}$ \\
conformal scalar & $-\frac{1}{360}$ & $\frac{1}{240}$ \\
graviton & $\frac{23}{60}$ & $\frac{7}{40}$ \\\hline\hline
\end{tabular} 
  \caption{\label{tab:self-energy} Different particles contribution to the graviton self-energy at the one-loop level, $a_1$ and $a_2$ are coefficients in the graviton form factor $\Pi^{\mu\nu\rho\sigma}$ as shown in Eq.~\eqref{eq:form_factor}. }
\end{table}

\section{The kernel structure}
In order to numerically obtain solutions for $\alpha(p^2)$ and $\beta(p^2)$ we rotate the expressions of 
\equaref{eq:alpha_1} and \equaref{eq:beta_1} respectively to Euclidean space. 
We begin by performing the following  replacements, 
 $k^2 = - k_E^2$, $p^2 = - p_E^2$, $d^4k=id^4k_E$ which modify $\alpha(p^2)$ and $\beta(p^2)$ to take the following form
\begin{figure*}[t!]
\centering
\includegraphics[width=0.85\textwidth]{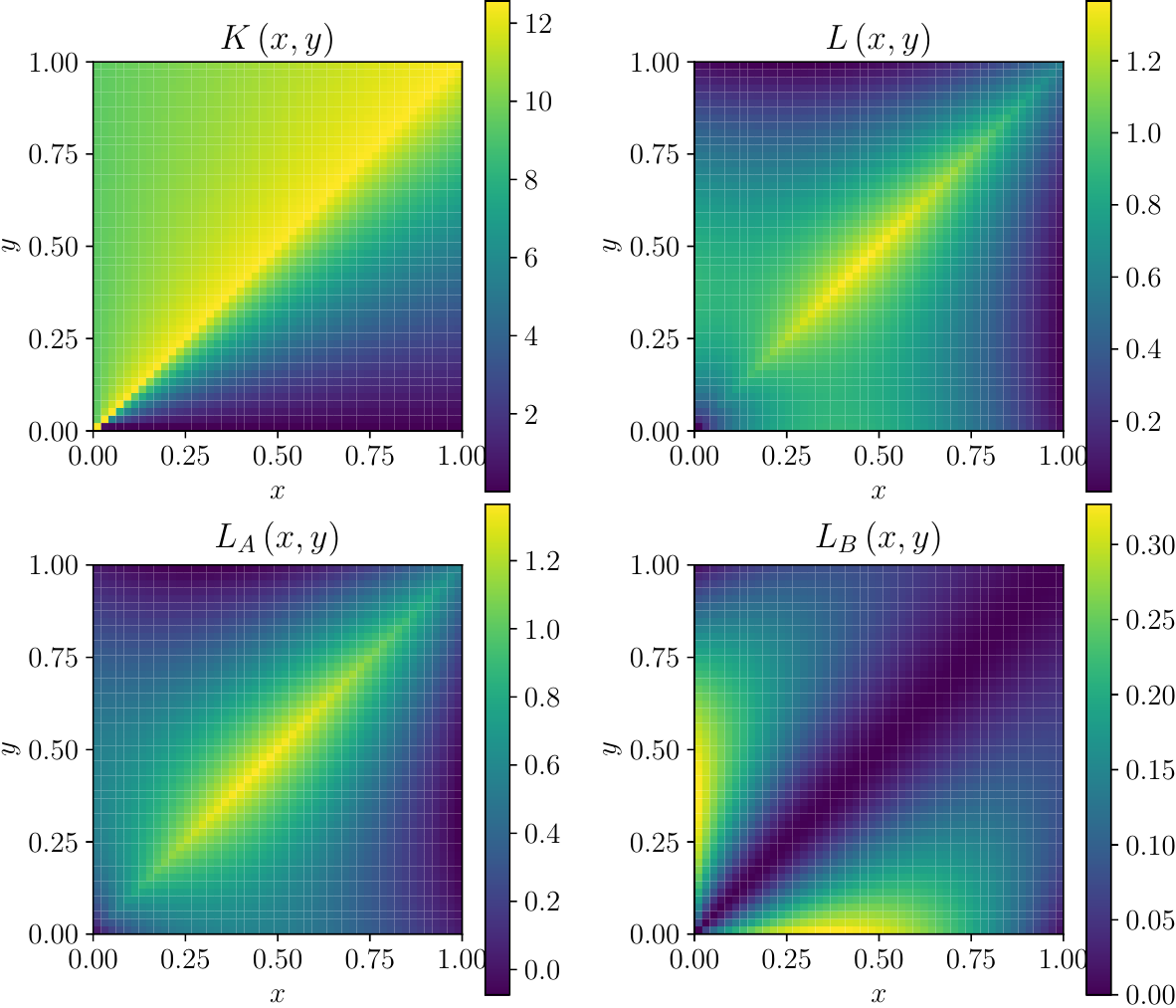}
\caption{For the SM particle content we display the kernels $K(x,y)$ and $L(x,y)= A L_{A}(x,y)+B L_{B}(x,y)$ for $x\in[0,1]$ and $y\in[0,1]$.}\label{fig:kernels}
\end{figure*}
\begin{equation} \label{eq:alpha_beta2}
\begin{aligned}
&\alpha(p_E^2) = 1 -2\pi G\\
 & \int[dk_{E}]f(k_E)
\frac{ \left[ 2(k_E \cdot p_E)^2+4k_E^2 p_E^2+3k_E \cdot p_E(k_E^2+p_E^2) \right]  }{p_E^2(p_E-k_E)^2}
 \,, \\
&\beta(p_E^2) = - 8 G^2\!\int[dk_{E}]f(k_E)\\
&\left[ A (k_E+p_E)^2\! - \!B \frac{\left(p_E^2-k_E^2\right)^2}{8 (p_E-k_E)^2} \right] \log \left[\frac{\mu^2}{(p_E-k_E)^2} \right] \,, \\
\end{aligned}
\end{equation}
where  $k_E^2$ and $p_E^2$ are  positive and the Euclidean rotation has changed the relative sign between $\alpha^2$ and $\beta^2$ in  the denominator of the above expression.
   We note that we have not calculated $\alpha(p_E^2)$ using the dressed graviton propagator because  its contribution will receive further
suppression, by a loop factor, than the leading non-zero undressed contribution. The most straightforward regularisation procedure, which is the one we adopt, is to impose an ultraviolet (UV) cutoff $\Lambda$ on the magnitude of the momentum running in the loop.  This UV cutoff is also the condensate scale, above which the chiral symmetry of the neutrino is restored and the condensate dissolves. We expect the UV cutoff to be the same order as $\mu$ given the non-renormalisability of quantum gravity.
In QED, a cutoff regularisation scheme is often employed due to its numerical convenience. However, such an approach has its disadvantages as it lacks Lorentz covariance and may lead to ambiguous results. As it has been demonstrated the cutoff regularisation scheme yields qualitatively similar solutions as those derived using off-shell renormalisation \cite{Hawes:1994ce,Hawes:1996mw}, we proceed with this approach.

 We rescale the momentum and $\beta$ such that $p_E^2 = x \Lambda^2$, $k_E^2 = y \Lambda^2$ and $\beta \to \beta\Lambda$. In addition,
 we replace the integral measure, $d^4 k_E$, by  the hyper-spherical
coordinates, $d^4 k_E = 2\pi \Lambda^4 y dy \sin^2\theta d\theta$,
  as well as defining $p_E\cdot k_E = \sqrt{xy} \Lambda^2 \cos\theta$ for $\theta \in [0, 2\pi]$ to  obtain 
\begin{equation} \label{eq:alpha_beta_2}
\begin{aligned}
\alpha(x) &= 1 - \frac{G\Lambda^2}{(2\pi)^2} \int_0^1 dy
\frac{y \alpha(y)}{ y \alpha^2(x) + \beta^2(y)} K(x,y)
 \,,\\
\beta(x) &= +\frac{8 G^2 \Lambda^4}{(2\pi)^3} \int_0^1 dy 
\frac{y \beta(y)}{y \alpha^2(y) + \beta^2(y)} L(x,y)\,.
\end{aligned}
\end{equation}

The variable transformed  kernels of \equaref{eq:alpha_beta2} may be written as 

\begin{equation} \label{eq:KL_integration}
\begin{aligned}
K(x,y) &= \frac{1}{x} \int_{0}^{\pi} s^2_{\theta} d\theta 
\frac{ 2 xy\cos^2\theta+4 xy+3\sqrt{xy}(x+y)c_{\theta}}{x+y-2\sqrt{xy}c_{\theta}} \,, \\
L(x,y) &= \int_{0}^{\pi}  s^2_{\theta} d\theta A \, (x+y+2\sqrt{xy}c_{\theta}) \log \left[x+y-2\sqrt{xy}c_{\theta} \right] \\
&- B \, \frac{(x-y)^2}{8 (x+y-2\sqrt{xy}c_{\theta})} \log \left[x+y-2\sqrt{xy}c_{\theta} \right] \,,
\end{aligned}
\end{equation}
where $ s_{\theta},  c_{\theta}$ are $\sin\theta$ and $\cos\theta$.
The integrated forms of the above equations shown in \figref{fig:kernels} for SM values of $A$ and $B$.
In the top left plot of \figref{fig:kernels} is the kernel of  $\alpha$, $K(x,y)$. This kernel is large, $K\approx\mathcal{O}(10)$,
for $x\sim y$. Because $\alpha$ is non-zero at leading order 
 it is not sensitive to the matter content and is therefore independent of parameters $A$ and $B$. 
Naturally, this is not the case for the kernel of $\beta$, $L(x,y)$, which is calculated using the dressed graviton propagator. For the same reasoning, the coefficient of the integrals of $\alpha$ and $\beta$ 
(given by $G\Lambda^2$ and $(G\Lambda^2)^2$ respectively) enter with different powers of the UV cutoff. 
The kernel $L(x,y)$ is split into two components, $L_{A}(x,y)$ and $L_{B}(x,y)$, which are premultiplied
by $A$ and $B$ respectively: 
\begin{equation}
L(x,y) = A L_{A}(x,y) + B L_B(x,y) \,. 
\end{equation}
$L_{A}(x,y)$ and $L_{B}(x,y)$ are shown in the bottom left and  right plots of  \figref{fig:kernels}.  
We observe that  $L_{A}(x,y)$ is  larger by a factor of a few than $L_B(x,y)$ for the majority of
the $x-y$ region and hence  the combined kernel $L(x,y)$ is  dominated by $L_{A}(x,y)$.
For $x \geqslant y$, the kernels have the simplified form:
\begin{equation} \label{eq:kerneltransform1}
\begin{aligned}
K(x,y) &= \frac{\pi}{x} \frac{y}{x} (3x+y) \,, \\
L_A(x,y) &= \frac{\pi}{12} \left[\frac{5y^2 - 3 xy}{x}  - 6 (x+y) \log x \right]\,,  \\
L_B(x,y) &=\frac{\pi}{8} \frac{(x-y)^2}{xy}
\left[ y - x \log x + (x-y)\log(x-y) \right]\,. 
\end{aligned}
\end{equation}
From \equaref{eq:KL_integration}, we observe  the $xK(x,y)$ and $L_{A,B}(x,y)$ are symmetric functions of $x$ and $y$. Therefore, we perform  the convenient replacement $x\to \frac{(x+y)+|x-y|}{2}$ and $y\to \frac{(x+y)-|x-y|}{2}$ into expressions of $xK(x,y)$ and $L_{A,B}(x,y)$ to  obtain 

\begin{equation}\label{eq:kerneltransform2}
\begin{split}
K(x,y) &= \frac{\pi}{x} \frac{(x+y)^3 - [(x+y)^2+2xy]|x-y| }{2xy} \\
L_A(x,y) &= \frac{\pi}{12} \left\{ \frac{5(x^2+y^2)-5(x+y)|x-y| - 6 xy}{(x+y)+|x-y|}  \right.\\
& \left. - 6 (x+y) \log \left[ \frac{(x+y)+|x-y|}{2} \right] \right\} \\
L_B(x,y) & = 
\frac{\pi}{8} \frac{(x-y)^2}{xy}\left\{\frac{(x+y)-|x-y|}{2}  \right.\\
&  \left.- \frac{(x+y)+|x-y|}{2} \log \left[ \frac{(x+y)+|x-y|}{2} \right]\right.\\
&\left.+ |x-y|\log(|x-y|) \right\}\,,
\end{split}
\end{equation}

which are expression for the kernels of $\alpha$ and $\beta$ respectively  if $x>y$ or $x<y$. 
The analytic manipulation from \equaref{eq:kerneltransform1} to  \equaref{eq:kerneltransform2} is applied to make the kernels more amenable for
numerical integration.

The non-trivial momentum structure of the $L(x,y)$ kernel may be most easily understood in the language of BCS theory \cite{Bardeen:1957mv} which describes the pairing of fermions.  A Cooper pair can form between two  fermions of opposite momenta and spin. In such a configuration, the system's energy is minimised, and the fermions combine to give a  spin-singlet (or possibly triplet), which leads to an attractive interaction between the fermionic pair. In addition to the spin component, there is also an orbital angular momentum component, $l$, which takes integer values; in the case of an $s$-wave interaction, $l=0$. However, the orbital components can also have other non-trivial integer values,  $l=1$ ($p$), $l=2$ ($d$) which characterise the pairing. As observed from the kernel structure shown in \figref{fig:kernels}, $L(x,y)=0$ for $x=y=0$ which corresponds to a $d$-wave interaction. This feature arises due to the spin-2 nature of the graviton.

\section{Numerical solutions to the Schwinger-Dyson equation}
It was postulated in \cite{Dvali:2016uhn} that active neutrinos could condense via gravitational instantons and consequently acquire a mass below energies of $\sim 100$ meV.  In this work, we apply the calculational techniques used to
condense right-handed neutrinos from their gravitational interactions \cite{Barenboim:2010db} to light, active neutrinos.

The practical challenge of this task comes from the great separation in the relevant energy scales. The gravitational coupling is parametrised by
$\kappa = \sqrt{32\pi G}\approx 10^{-18}\, \text{GeV}^{-1}$.  From the SDE we find the coefficients of both $\alpha$ and $\beta$
 are proportionate to $G\Lambda^2$ and $(G\Lambda^2)^2$, respectively. To recover a non-trivial vacuum, 
 the cutoff scale $\Lambda$ cannot be far from the Planck scale ($M_{\rm pl}\approx 10^{19}$ GeV) unless the particle content 
 is enormous. This point will become more apparent in \secref{sec:extra}. On the other end of the energy scale
are the tiny neutrino masses,  $m_{\nu}\sim 10^{-10}$ GeV.\footnote{It was very recently confirmed  by the terrestrial experiment KATRIN that the effective neutrino mass measured using beta decay is less than $1.1$ eV ($90\%$ CL) \cite{Aker:2019uuj}.} 
We can contrast this with the relevant scales in QCD, where $\Lambda_{\text{QCD}}\approx 1$ GeV and the light quark masses, $m_{u/d} \approx \mathcal{O}(1)$ MeV. We find the separation in scales to be $m_{u/d}/\Lambda_{\text{QCD}}\approx 10^{-3}$. However, in our case of interest, 
we require a non-zero but very small ratio of scales, $m_{\nu}/\Lambda_{G}\approx 10^{-29}$. 
This presents a unique numerical challenge in finding the non-trivial vacuum compared with
other gauge theories where the SDE techniques are applied.
We present two possible solutions below which demonstrate gravitationally induced chiral symmetry breaking.
\subsection{Extrapolation}\label{sec:extra}
   \begin{figure}[t!]
\centering
\includegraphics[width=0.75\textwidth]{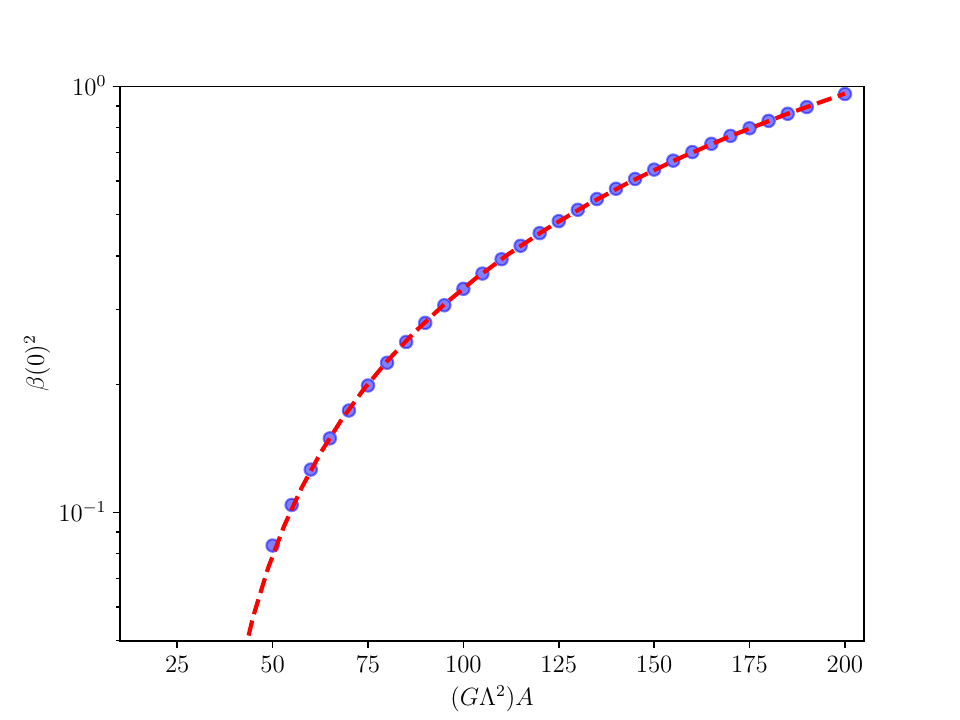}
\caption{$\beta^2(0)$ as  a function of $(G \Lambda^2)A$ where $A$ and $B$.  We take $A=B$ with $G \Lambda^2=1$ and vary the values of $A$ . The blue dots are the solutions to the SDE and the red dotted line is the fitted polynomial.}\label{fig:extrapolate}
\end{figure}
We implement  an  iterative, numerical method of solving the SDE with a cutoff regularisation a similar method as outlined in \cite{Abe:1983in}. The limitations such an 
approach are discussed in \cite{Abe:1984vk}.
We define $\alpha^{(i+1)}$ and $\beta^{(i+1)}$ (for $i=0,1,2,3,\cdots$)
 to be
\begin{equation}\label{eq:SDEit}
\begin{aligned}
\alpha^{(i+1)}(x) &= 1 - \frac{G \Lambda^2}{(2\pi)^2} \int_0^1 dy
\frac{y \alpha^{(i)}(y)}{ y {\alpha^{(i)}}^2(x) + {\beta^{(i)}}^2(y)} K(x,y)
 \,,\\
\beta^{(i+1)}(x) &= \frac{8(G \Lambda^2)^2}{(2\pi)^3} \int_0^1 dy 
\frac{y {\beta^{(i)}}(y)}{y {\alpha^{(i)}}^2(y) + {\beta^{(i)}}^2(y)} L(x,y) \,.
\end{aligned}
\end{equation}
We choose two trial functions as initial values for the iterative calculation
\begin{equation}
\alpha^{(0)}(x) = c_{1} \,, \quad \beta^{(0)}(x) = c_{2} \,,
\end{equation}
where $c_{1}$ and $c_{2}$ are constants. As expected, the solutions of  
 $\alpha$ and $\beta$ 
 do not exhibit sensitivity to the value of the trial functions.
 Also, we require a definition of convergence which we parametrise
 by the \emph{tolerance}
 \begin{equation}
 \text{tolerance} \equiv \frac{\beta^{(i+1)}(x)}{\beta^{(i)}(x)} - 1 \,.
 \end{equation}
 
 The procedure for solving \equaref{eq:SDEit} is as follows:
\begin{enumerate}
\item Choose a value of $G \Lambda^2$, $A$, $B$, tolerance and trial function values.
\item Subdivide the $x$-interval $[x_{\text{IR}},1]$ into $n$ bins, where $x_{\text{IR}}$ is 
infrared boundary of the theory. 
\item Iteratively solve \equaref{eq:SDEit} for each bin. 
\item For each bin we calculate the tolerance and summate this measure over all bins. 
\item Require the solution to be stable as the tolerance is reduced. 
\end{enumerate}
To ensure the solution is independent of the number of bins, we normalise the tolerance by the total number of bins. Additionally, 
we  test the solution does not vary for differing values of the tolerance.  As the region of interest is in far-infrared,
as represented by $x_{\text{IR}}$, to probe the
sub-electroweak energy scales requires $x\approx 10^{-34}$. Moreover, as the mass of the neutrino 
is defined as 
 \begin{equation}\label{eq:mass2}
m_{\nu} =\frac{\beta(0)}{\alpha(0)}\Lambda \,,
 \end{equation}
  requires $\beta(0)\approx 10^{-29}$ for $\Lambda\approx M_{\rm pl}$.
Achieving this level of precision in the numerical integration and the endpoint  is challenging, and consequently, we solve the 
 iterative SDE in regions where we have numerical control and then extrapolate in the combination $G \Lambda^2 A$. 
 
 In summary, fix $G \Lambda^2=1$ and vary  $A$ and  $B$ (which parametrise the matter content) and find the   non-trivial minima which gives rise to a non-zero stable value of  $\beta(x_{\text{IR}} = 10^{-10})$. We then extrapolate this function to $x_{\text{IR}}=0$ for $\beta$ to $\beta(x=0)$. We repeat this procedure for several values  $A$ and $B$ (for a fixed $G \Lambda^2$) and calculate the solutions numerically. We fit a polynomial to these points as represented by the red line and blue dots in \figref{fig:extrapolate} respectively.  The functional form of this polynomial is
  \begin{equation}\label{eq:extragrad}
  \begin{aligned}
 \beta^2(0)  &= -0.0792286 + 5.28\times 10^{-4} (G\Lambda^2A)^{\frac{3}{2}}\\
 & - 1.14\times 10^{-5}  (G\Lambda^2A)^2
 \end{aligned}
 \end{equation}
 This is the lowest order polynomial that provides a good fit to the numerically calculated points. To recover neutrino masses of the 
 correct order of magnitude requires $ \beta(0)\approx 10^{-29}$ with $G\Lambda^2\approx 1$  implies $A\gtrapprox 30$.
 Therefore new particle content is required to support the condensate and the lower the condensation scale the larger the particle content  required for chiral symmetry breaking to occur.  As instantons are a modification to the gauge boson propagator, the inclusion 
 of such effects are unlikely to lower the scale substantially. 
 We note that there is a large amount of fine-tuning required to reproduce light neutrinos masses: 
we must tune the quantity $G \Lambda^2A$  such that the solution is very close to but not equal to the chiral preserving solution,  $\beta(0)=0$.
This tuning is not surprising: in the scenario of minimal new particle content, the neutrino mass  is proportionate to  $\Lambda$, the only dimensionful parameter of \equaref{eq:mass2}. A non-trivial vacuum requires $G\Lambda^2 \approx 1$ otherwise the iterative solution of \equaref{eq:SDEit} evolves to the trivial vacuum.  Because $G\approx 1/M^2_{\rm pl}$ this implies  $\Lambda\approx M_{\rm pl}$. Therefore to recover sub-eV masses of neutrinos requires $\beta(0)$ to be very small. It is important to note that this discussion applies only to a minimal number of new degrees of freedom and the conventional scale of quantum gravity. However, as we will further elucidate in \secref{sec:discussion}, further new physics may ameliorate this fine-tuning.

 \begin{figure}[t!]
\centering
\includegraphics[width=0.6\textwidth]{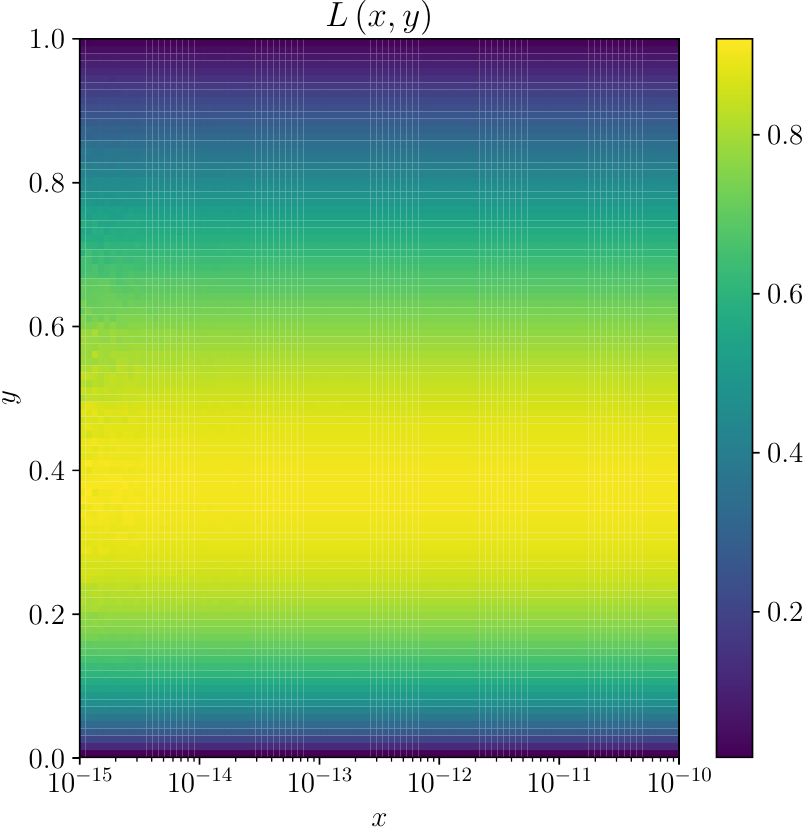}
\caption{For the SM particle content we display the kernels $L(x,y)=A L_{A}(x,y)+B L_{B}(x,y)$ for $x\in [10^{-15},10^{-10}]$ and $y\in[0,1]$.}\label{fig:kernelslog}
\end{figure}


\subsection{Consistency of $\beta(x)$ with the Kernel structure}\label{sec:const}
In this method, we exploit the structure of the kernel of $\beta$ to demonstrate the existence of a non-trivial vacuum. Once the true vacuum is found, \emph{any value} of the infrared mass can be recovered from parameter tuning.

\begin{figure}[t!]
\centering
\includegraphics[width=0.75\textwidth]{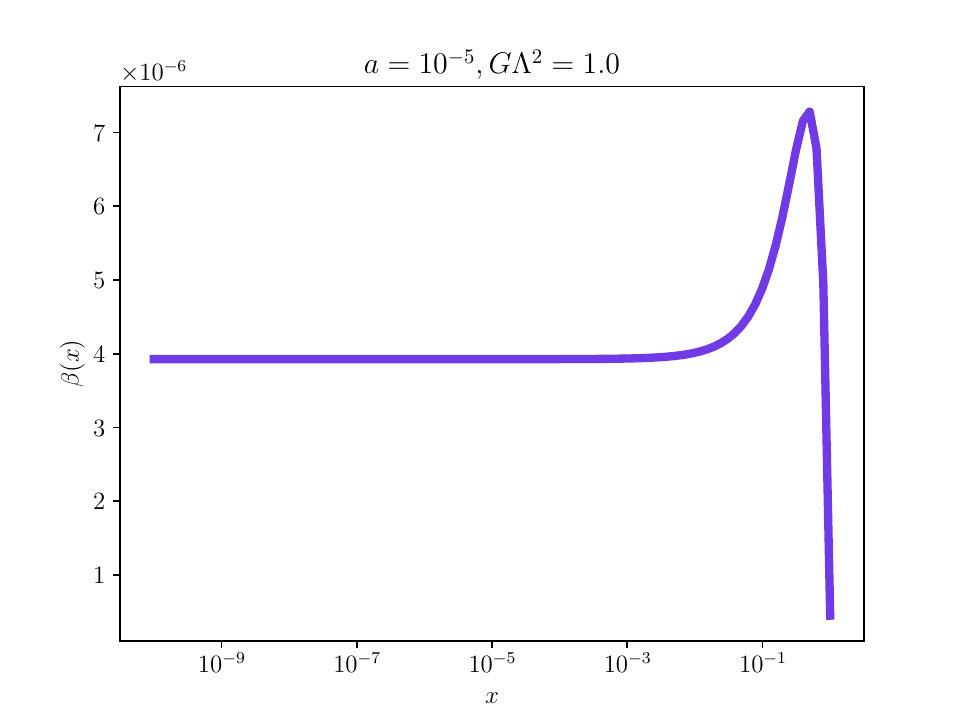}
\caption{$\beta(x)$ as a function of $x$ from solving \equaref{eq:approx1} with $a=10^{-5}$ and $G\Lambda^2=1$.}\label{fig:betaconst}
\end{figure}

 We remind the reader $x$ is the neutrino's ingoing momentum normalised by the UV cutoff and $y$ is the cutoff normalised loop momentum. 
From  \figref{fig:kernelslog}, we show the kernel of $\beta$, $L(x,y)$, for all values of $y$ and $x\in[10^{-15},10^{-10}]$
 which demonstrates the far-infrared behaviour of the kernel.\footnote{We note that for smaller values of $x$ ($x\ll10^{-15}$) this flat behaviour in $x$ does not change.} 
 From the kernel structure, we postulate that  $\beta(x)$ is insensitive to $x$  until close to the ultraviolet cutoff. 
 As the only $x$ dependence can come from the kernel because $y$  is integrated over, taking the form of $\beta$ to constant in $x$ is beyond a sufficiently good approximation. The same is true for the kernel of $\alpha$. 
From these considerations, the crudest approximation of $\beta(x)$ is a step function of  constant magnitude, $a$, which can be written as 
\begin{equation}\label{eq:approx1}
\beta(x) = \frac{8 (G\Lambda^2)^2}{(2\pi)^3} \int_{0}^1 \frac{a y dy}{a^2+y} L_A(x,y) \,,
\end{equation}
where  the quenched limit has been applied, $\alpha\approx 1$, and  the definition of $a$ contains information about the particle content and the overall scale of $\beta$.  Since $A>B$ and $L_{A}(x,y)>L_B(x,y)$ we have ignored the sub-leading contribution of $B L_{B}(x,y)$.  We solve  \equaref{eq:approx1} for a fixed value of $G\Lambda^2$ and $a$. For $a=10^{-5}$ and $G\Lambda^2 = 1$ we find the solution to be $\beta(0)\approx 4\times 10^{-6}$ as shown in \figref{fig:betaconst}.  With the appropriate tuning of $G\Lambda^2$ and $a$ any non-zero value of $\beta(0)$, and therefore the infrared mass of the neutrino, can be recovered.  From \figref{fig:betaconst}, we observe there is some  non-trivial $x$ dependence in $\beta(x)$. However, this only occurs for $x\approx 0.4$ which is in the far ultraviolet region of the theory.
As we are only interested in the  infrared mass and do not make any statements regarding the 
deep UV physics,  this feature does not impact the final result. 
This approach checks for the self-consistency of the postulated form
 of $\beta(x)$ with the kernel structure and agrees with the first method discussed in \secref{sec:extra}.

\section{Discussion}\label{sec:discussion}

In \secrefs{sec:extra}{sec:const} we demonstrated a common neutrino mass scale can be 
generated  through iteratively solving the SDE and also making an informed Ansatz for the 
form of $\beta(x)$.  From the first method, we found there are two factors  which support a 
gravitationally induced neutrino condensate: the scale of the condensate, $\Lambda$, and the particle content as parametrised by $A$ and $B$. These two factors compensate for each other: if the condensation scale is lowered, the particle content must be increased and vice versa.  The minimal particle content required for $\Lambda\approx M_{\rm pl}$ is still larger than the SM; however, this conclusion follows from assuming a conventional Planck scale, $M_{\rm pl}\approx 10^{19}$ GeV.  

An alternative  possibility comes from extra compact dimensions \cite{ArkaniHamed:1998rs, Randall:1999ee}. Given  $n$ extra dimensions, with length scale $R$, the number of degrees of freedom for each bulk field $N_p$ is proportional to the number of Kaluza-Klein (KK) excitations, and the latter is determined to the cutoff scale, i.e., $N_p \sim (\Lambda R)^n$ \cite{ArkaniHamed:1998rs}.
In this framework, the cutoff scale, $\Lambda$, should be lower than the true Planck scale $M_*$ in $4+n$ dimensions, which is correlated with the effective Planck scale $M_{\rm pl}$ in the four dimensions. For example, in the ADD model \cite{ArkaniHamed:1998rs}, $M_{\rm pl}^2 = M_*^{n+2} R^n$, $N_p$ can maximally reach the order $(M_* / M_{\rm pl})^2$. 
The true Planck $M_*$ is much lower than the effective scale $M_{\rm pl}$. By assuming the cutoff scale to be  just below $M_*$, e.g., $\Lambda = 0.9 M_*$, we obtain $G\Lambda^2A \sim \Lambda^2 N_p / M_{\rm pl}^2 \sim \mathcal{O}(1)$, which guarantees the model staying in the non-perturbative regime. In this regime, the ratio between neutrino mass and the scale is modified into $m_\nu / \Lambda \sim m_\nu / M_*$.  
Furthermore, by taking $n=3$ and $R\sim 10^{-9} $ m, which is sufficient to evade experimental constraints \cite{Csaki:2018muy}, $M_*$ is lowered to the TeV scale and $m_\nu / \Lambda \sim 10^{-14}$. 

In contrast with the vast majority of neutrino mass generation mechanisms, the effect of gravitational condensation is insensitive to the Dirac or Majorana nature of the neutrino. 
As gravity is universal and does not discriminates between particle species, if a small mass is induced for neutrinos, a small mass will be induced for the other SM fermions at the scale $\Lambda$. 
Although this scale may be significantly below the Planck scale, a great deal of new matter is required. We note that electroweak symmetry will not be induced as the effect of tadpoling the Higgs
is of order of the mass contribution to the neutrinos, which is small.  The connection between  the gravity induced mass gap for the neutrinos and black holes, which are non-perturbative solutions in 
general relativity, may be recovered  as poles in the resummed graviton propagator could be interpreted as black hole precursors \cite{Calmet:2014gya,Calmet:2015pea}.

In summary, we have shown that neutrinos can condense via gravitational interactions and undergo chiral symmetry breaking.  To do so, we treat gravity as an effective quantum field theory
and solve the Schwinger-Dyson equations to find a non-trivial vacuum. The true vacuum is recovered in two ways: the first through iteratively solving the SDE and the second from making an  Ansatz for the kernel of $\beta$. In the minimal setup, the scale of the condensation is found to be close to the Planck scale and new degrees of freedom beyond the Standard Model particle content are required.
 Interestingly, new physics is required to explain neutrino mass scale in this framework:
 the Standard Model in addition to gravity is insufficient to explain neutrino masses.  An important point to note is that this calculation demonstrates a 
  common neutrino mass scale may be gravitationally induced; however,
 to reproduce oscillation data, a further mechanism is required to break the mass degeneracy.
 We agree with the conclusions of \cite{Dvali:2016uhn}, where a common mass scale is recovered. 
 In that work, the neutrinos' mass splittings are induced from the effective potential of the Goldstone bosons, which is a non-gravitational effect. The Goldstone bosons
 associated with the former mechanism  can be relatively light ($\sim \,\rm{MeV}$)  and there is interesting associated phenomenology  \cite{Funcke:2019grs}. However, 
 in this work, if the condensate scale is high (and not lowered due to extra-dimensions or new particle content), the mass of the Goldstone bosons would be large and no such low-scale phenomenology would not be observed.
 On the other hand, if the scale was lowered  due to extra-dimensions or new particle content, similar phenomenology to that discussed in  \cite{Funcke:2019grs} would be observed. Furthermore, lowering the condensate scale would have interesting cosmological consequences: as the neutrino mass changes on cosmological timescales, its mass can be reconstructed as a function of redshift \cite{Lorenz:2021alz}.

 We find that with minimal new particle content, the condensation scale is high and close to the Planck scale. The contribution of the condensate to neutrino masses is small and 
 equal for the three neutrino masses. The condensate scale can be lowered if there is a significant increase in the number of degrees of freedom and in such a case, the contribution of the condensate to the three
 neutrino masses would still be small and equal. This compensatory effect in supporting the condensate does not fully overcome the large fine-tuning required to recover tiny neutrino masses.  However, if the true Planck scale is lowered than expected, due to extra compact dimensions, such tuning could be somewhat reduced.

\section*{acknowledgments}
We are grateful to Craig Roberts for generously sharing his knowledge of the Schwinger Dyson equations and non-perturbative physics with us.
It is a pleasure to thank Simon Badger, Bill Bardeen, Nikita Blinov and Pedro Machado for  useful discussions on various aspects of this work. 
We thank Niels Bjerrum-Bohr and Apostolos Pilaftsis for their correspondence regarding the graviton-fermion-fermion Feynman rules. We are grateful to Joannis Papavassiliou for useful discussion on non-perturbative techniques. GB acknowledges support from the MEC and FEDER (EC) Grant SEV-2014-0398, FIS2015-72245-EXP, and FPA2017-845438 and the Generalitat Valenciana under grant PROMETEOII/2017/033 and also partial support  from the  European  Union  FP10 ITN ELUSIVES (H2020-MSCAITN-2015-674896) and INVISIBLES-PLUS (H2020-MSCA-RISE-2015-690575). This manuscript has been authored by Fermi Research Alliance, LLC under Contract No. DE-AC02-07CH11359 with the U.S. Department of Energy, Office of Science, Office of High Energy Physics.  YLZ acknowledges the STFC Consolidated Grant ST/L000296/1 and the European Union's Horizon 2020 Research and Innovation Programme under Marie Sk\l{}odowska-Curie grant agreements Elusives ITN No.\ 674896 and InvisiblesPlus RISE No.\ 690575. JT and YLZ also gratefully acknowledge the hospitality of Universitat de Valencia.


\appendix 

\section{Graviton Feynman rules}\label{sec:FR}
We follow the convention of  \cite{BjerrumBohr:2004mz,Donoghue:2017pgk} and present some of the basics of gravitational field theory in this Appendix. The full gravitational action is given
by 
\begin{equation}
S_g = \int d^4 x \sqrt{-g} (\frac{1}{4\pi G} R + \mathcal{L}_m) \,,
\end{equation}
where $R$ is the scalar curvature and $\mathcal{L}_m$ is the Lagrange density for matter. For massless minimal scalar (spin-0), Dirac fermion (spin-1/2), and gauge boson (spin-1) particles, as denoted by $\phi$, $\psi$ and $A_\mu$ respectively, the $\mathcal{L}_m$ term are represented by 
\begin{equation}
\begin{aligned}
\mathcal{L}_m &= D_\mu \phi^* g^{\mu\nu} D_{\nu} \phi + 
 \frac{i}{2} \left[ \bar{\psi} \gamma^a e_a^\mu D_{\mu} \psi + (D_\mu\bar{\psi}) \gamma^a e_a^\mu \psi \right] \\
 &- \frac{1}{4}
 g^{\mu\nu} g^{\rho \sigma}F_{\mu\rho} F_{\nu\sigma} \,, 
 \end{aligned}
\end{equation}
where $F_{\mu\nu} = D_{\mu} A_\nu - D_{\nu} A_\mu$ and $D_\mu$ denotes the covariant derivative with respect to the gravitational field and gauge fields, and $e_a^\mu$ is the vierbein to shift frame to the local Minkowski flat frame. 
In the flat space background, Feynman rules for gravitational interactions are obtain by perturbing the metric 
\begin{equation}
g_{\mu\nu}  \rightarrow \eta_{\mu\nu} + \kappa h_{\mu\nu} \,, 
\end{equation}
where $\kappa = \sqrt{32\pi G}$. 
As  we work in the flat space background, the classical gravitational field is fixed at zero and $h_{\mu\nu}$ represents the gravitational quantum perturbation. 
The tree-level Feynman rules for gravitation propagator is given by
\begin{equation}
\text{Graviton propagator}:  G_{\mu\nu\rho\sigma}(p) = \frac{i \mathcal{P}_{\mu\nu\rho\sigma}}{p^2}
\end{equation}
with 
\begin{equation}
\mathcal{P}^{\mu\nu\rho\sigma} = \frac{1}{2}(\eta^{\mu\rho} \eta^{\nu\sigma} + \eta^{\mu\sigma} \eta^{\nu\rho} - \eta^{\mu\nu} \eta^{\rho \sigma}) \,.
\end{equation}
The tree-level Feynman rules for massless minimal scalar, Dirac fermion and gauge boson propagators are respectively given by
\begin{equation}
\begin{aligned}
\text{Minimal scalar propagator}:  \Delta(p)& = \frac{i}{p^2} \,,\\
\text{Dirac fermion propagator}: S_F(p) &= \frac{i}{p\!\!\!\slash} \,, \\
\text{Gauge boson propagator}:  D_{\mu\nu}(p)& = \frac{i \eta_{\mu\nu}}{p^2} \,. 
\end{aligned}
\end{equation}

Feynman rules for interactions between graviton and fermions, minimal scalars and gauge bosons are 
given by $\tau_{1}$, $\tau_{2}$ and $\tau_{3}$ respectively. \\[5mm]

\begin{equation}
\begin{aligned}
&{\hbox{\includegraphics[width=0.2\textwidth]{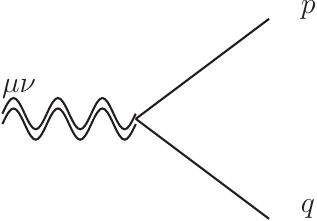}}}\\
&\tau^{\mu\nu}_1(p,q) =\frac{ i\kappa }{8} \left[ (q-p)^\mu \gamma^\nu + (q-p)^\nu \gamma^\mu - 2\eta^{\mu\nu} (q\!\!\!\slash - p\!\!\!\slash)
 \right]
 \end{aligned}
\end{equation}

\begin{equation}
\begin{aligned}
&{\hbox{\includegraphics[width=0.2\textwidth]{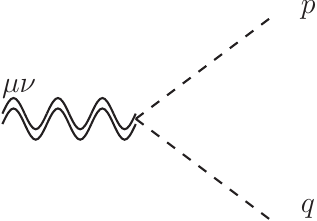}}}\\
&\tau^{\mu\nu}_1(p,q) =\frac{ i\kappa }{8} \left[ (q-p)^\mu \gamma^\nu + (q-p)^\nu \gamma^\mu - 2\eta^{\mu\nu} (q\!\!\!\slash - p\!\!\!\slash)
 \right]
  \end{aligned}
\end{equation}

\begin{equation}
\begin{aligned}
&{\hbox{\includegraphics[width=0.2\textwidth]{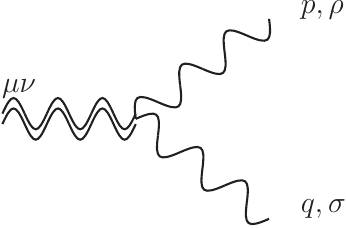}}}\\
&\tau^{\mu\nu\rho\sigma}_3(p,q) = i\kappa [-\mathcal{P}^{\mu\nu\rho\sigma} - \frac{1}{2} \eta^{\mu\nu} p^\sigma q^\rho + \eta^{\sigma \rho}(p^\mu q^\nu +p^\nu q^\mu)& \nonumber\\
& \hspace{1cm} +\frac{1}{2} ( \eta^{\mu\sigma} p^\nu q^\rho + \eta^{\nu\sigma} p^\rho q^\mu + \eta^{\nu\rho} p^\sigma q^\mu + \eta^{\mu\rho} p^\nu q^\sigma)] 
 \end{aligned}
\end{equation}
  where $\kappa = \sqrt{32\pi G}$ and  all momenta are assumed to be out-flowing from the vertex.


\section{Vacuum polarisation calculation}\label{sec:vacpolapp}%
In this section we show the brute force method of calculating the vacuum polarisations. Lorentz algebra was 
manipulated using FeynCalc \cite{Shtabovenko:2016sxi} and  the loop integration was completed using Package X \cite{Patel:2015tea}.
The contribution from gauge bosons  to the vacuum polarisation tensor is
\begin{equation}
\begin{aligned}
i\Pi^{\alpha\beta,\gamma\delta} &= \int[dl]{\tau^{\alpha\beta\rho\sigma}_{3}}(l,q-l)\frac{-i  g_{\rho\sigma}(l+q)^2}{(l+q)^4}\\
&\tau^{\gamma\delta\xi\omega}_{3}(-l,l-q)\frac{-i g_{\xi\omega}l^2}{(l)^4} \\
 &= \kappa^2  \int[dl]\tau_{3}^{\alpha\beta\rho\sigma}(l,q-l)\frac{g_{\rho\sigma}(l+q)^2}{(l+q)^4}\\
 &\tau_{3}^{\gamma\delta\xi\omega}(-l,l-q)\frac{g_{\xi\omega}l^2}{(l)^4} \\
& =\frac{2iG}{\pi}\big[
\frac{1}{30}\left( q^{\alpha} q^{\beta}-q^2g^{\alpha\beta} \right)\left( q^{\gamma} q^{\delta}-q^2g^{\gamma\delta} \right)\\
&-\frac{1}{20}\left( q^{\alpha} q^{\gamma}-q^2 g^{\alpha\gamma} \right)\left( q^{\beta} q^{\delta}-q^2 g^{\beta\delta} \right)\\
& -\frac{1}{20}\left( q^{\alpha} q^{\delta}  -q^2g^{\alpha\delta} \right)\left( q^{\beta} q^{\gamma}-q^2g^{\beta\gamma} \right)
\big]\log\left[-\frac{\mu^2}{q^2}\right].
\end{aligned}
\end{equation}
where we have divided by a symmetry factor of two.
The contribution from Dirac fermions to the vacuum polarisation tensor is
\begin{equation}
\begin{aligned}
i\Pi^{\alpha\beta,\gamma\delta} &= \kappa)^2\int[dl]\tau_1^{\alpha\beta}(l,q-l)\frac{(\slashed{l}+\slashed{q})}{(l+q)^2}\tau_1^{\gamma\delta}(-l,l-q)\frac{(\slashed{l})}{(l)^2} \\
&= i\frac{2G}{\pi}\big[
-\frac{2}{15}\left( q^{\alpha} q^{\beta}-q^2g^{\alpha\beta} \right)\left( q^{\gamma} q^{\delta}-q^2g^{\gamma\delta} \right)\\
&+\frac{1}{5}\left( q^{\alpha} q^{\gamma}-q^2 g^{\alpha\gamma} \right)\left( q^{\beta} q^{\delta}-q^2 g^{\beta\delta} \right) \\ 
&+\frac{1}{5}\left( q^{\alpha} q^{\delta}-q^2g^{\alpha\delta} \right)\left( q^{\beta} q^{\gamma}-q^2g^{\beta\gamma} \right)
\big]\log\left[-\frac{\mu^2}{q^2}\right].
\end{aligned}
\end{equation}
and the  contribution from minimal scalars to the vacuum polarisation tensor is
\begin{equation}
\begin{aligned}
i\Pi^{\alpha\beta,\gamma\delta} &= \int[dl]\tau_{2}^{\alpha\beta}(l,q-l)\frac{i}{(l+q)^2}\tau_{2}^{\gamma\delta}(l+q,l)\frac{i}{(l)^2} \\
 &=\kappa^2 \frac{i}{16\pi^2}  \int[dl]\frac{\tau_{2}^{\alpha\beta}(l,q-l)\tau_{2}^{\gamma\delta}(-l,l-q)}{(l+q)^2l^2} \\
&= i\frac{2G}{\pi}\big[
\frac{1}{40}\left( q^{\alpha} q^{\beta}-q^2g^{\alpha\beta} \right)\left( q^{\gamma} q^{\delta}-q^2g^{\gamma\delta} \right)\\
&+\frac{1}{240}\left( q^{\alpha} q^{\gamma}-q^2 g^{\alpha\gamma} \right)\left( q^{\beta} q^{\delta}-q^2 g^{\beta\delta} \right) \\ 
&+\frac{1}{240}\left( q^{\alpha} q^{\delta}-q^2g^{\alpha\delta} \right)\left( q^{\beta} q^{\gamma}-q^2g^{\beta\gamma} \right)
\big]\log\left[-\frac{\mu^2}{q^2}\right].
\end{aligned}
\end{equation}
Although the graviton is not a matter field, it does have self couplings and therefore the graviton will contribute to its own vacuum polarisation. We apply the contribution from the graviton self vacuum polarisation as calculated by 't Hooft  and Veltman \cite{tHooft:1974toh},
\begin{equation}
\begin{aligned}
i\Pi^{\alpha\beta,\gamma\delta}&= i\frac{2G}{\pi}\big[
\frac{23}{60}\left( q^{\alpha} q^{\beta}-q^2g^{\alpha\beta} \right)\left( q^{\gamma} q^{\delta}-q^2g^{\gamma\delta} \right)\\
&-\frac{7}{40}\left( q^{\alpha} q^{\gamma}-q^2 g^{\alpha\gamma} \right)\left( q^{\beta} q^{\delta}-q^2 g^{\beta\delta} \right) \\ 
&-\frac{7}{40}\left( q^{\alpha} q^{\delta}-q^2g^{\alpha\delta} \right)\left( q^{\beta} q^{\gamma}-q^2g^{\beta\gamma} \right)
\big]\log\left(-\frac{\mu^2}{q^2}\right).
\end{aligned}
\end{equation}

\bibliographystyle{unsrt}
\bibliography{grav}

\end{document}